\documentclass[twocolumn,prb,aps,10pt]{revtex4-1}

\usepackage{amsfonts,amssymb}
\usepackage[sumlimits,intlimits]{amsmath}
\usepackage{graphics}
\usepackage{graphicx}
\usepackage{mathrsfs}
\usepackage{wasysym}
\usepackage{textcomp}
\usepackage{verbatim}
\usepackage{hyperref}    
\usepackage{bm}          

\newcommand{\be}{\begin{equation}}
\newcommand{\ee}{\end{equation}}
\newcommand{\sgn}{\operatorname{sgn}}
\newcommand{\rr}{{\mathbf{r}}}
\newcommand{\RR}{{\mathbf{R}}}
\newcommand{\phit}{\widetilde{\phi}_1}
\newcommand{\smin}{\sigma_\text{min}}
\newcommand{\Tes}{T_\text{ES}}
\newcommand{\e}{\varepsilon}
\newcommand{\tdos}{\langle \nu \rangle}

\begin{document}

\title{Effects of bulk charged impurities on the bulk and surface transport in three-dimensional topological insulators\\
{\em \small Dedicated to the memory of Professor Anatoly Larkin} }

\author{Brian Skinner}
\author{Tianran Chen}
\author{B. I. Shklovskii}
\affiliation{Fine Theoretical Physics Institute, University of Minnesota, Minneapolis, MN 55455, USA}

\date{\today}

\begin{abstract}

In the three-dimensional topological insulator (TI), the physics of doped semiconductors exists literally side-by-side with the physics of ultra-relativistic Dirac fermions.  This unusual pairing creates a novel playground for studying the interplay between disorder and electronic transport.  In this mini-review we focus on the disorder caused by the three-dimensionally distributed charged impurities that are ubiquitous in TIs, and we outline the effects it has on both the bulk and surface transport in TIs.  We present self-consistent theories for Coulomb screening both in the bulk and at the surface, discuss the magnitude of the disorder potential in each case, and present results for the conductivity.  In the bulk, where the band gap leads to thermally activated transport, we show how disorder leads to a smaller-than-expected activation energy that gives way to VRH at low temperatures.  We confirm this enhanced conductivity with numerical simulations that also allow us to explore different degrees of impurity compensation.  For the surface, where the TI has gapless Dirac modes, we present a theory of disorder and screening of deep impurities, and we calculate the corresponding zero-temperature conductivity.  We also comment on the growth of the disorder potential as one moves from the surface of the TI into the bulk.  Finally, we discuss how the presence of a gap at the Dirac point, introduced by some source of time-reversal symmetry breaking, affects the disorder potential at the surface and the mid-gap density of states.

\end{abstract}
\maketitle

\section{Introduction}
\label{sec:Intro}


The three-dimensional (3D) topological insulator (TI)~\cite{Fu2007tii, Moore2007tio, Roy2009tpa, Fu2007tiw, Qi2008tft} has generated a great deal of excitement in the physics community because of its gapless surface states, which host a spectrum of quantum transport phenomena~\cite{Hasan2010c:t, Qi2011tia}.  Unfortunately, while a number of crystals have been identified to be 3D TIs, most of them are not actually insulators, but instead have a relatively large bulk conductivity that shunts the surface conductivity for TI crystals of substantial thickness ($ \gtrsim 10$\,$\mu $m).  How to achieve a bulk-insulating state is a problem that is widely discussed in the current literature~\cite{Qu2010qoa, Analytis2010tds, Checkelsky2009qii, Butch2010sss, Analytis2010bfs, Eto2010aoo, Ren2011oot, Ren2011o$s, Ren2012flt}. 

Typically, as-grown TI crystals are heavily doped $n$-type semiconductors, so that the Fermi level resides in the bulk conduction band.  In order to arrive at a bulk insulating state, such TIs are compensated by acceptors. With increasing compensation $K =N_A/N_D$, where $N_D$ and $N_A$ are the concentrations of monovalent donors and acceptors, respectively, the Fermi level shifts from the conduction band to inside the gap and then into the valence band. When compensation of donors is complete, $K=1$, the Fermi level is in the middle of the gap and the most insulating state of the TI is reached. The hope is that for a TI with bulk band gap $E_g \sim 0.3$ eV (as, for example, in Bi$_2$Se$_3$) the bulk resistivity should obey the activation law 
\be
\rho = \rho_0 \exp(\Delta/k_B T)
\label{act}
\ee
with activation energy $\Delta = E_g/2 \sim 0.15$\,eV, so that at room temperatures and below the TI is well insulating. 

The typical experimental situation near $K=1$, however, is  frustrating~\cite{Ren2011o$s}. In the range of temperatures between 100\,K and 300\,K the resistivity is activated, but with an activation energy that is roughly three times smaller than expected, $\Delta \sim 50$\,meV.  At $T \sim 100$\,K the activated transport is replaced by variable range hopping (VRH) and the resistivity grows even more slowly with decreasing $T$. Finally, at even smaller temperature, $T < 50$\,K, the resistivity saturates
\footnote{The authors of Ref.\ \onlinecite{Ren2011o$s} interpret this saturation as the contribution of the surface states.}
at a value $< 10$\,$\Omega$cm. 

In a recent paper~\cite{Skinner2012wib} we showed that the unexpectedly large bulk conductivity of TIs at $K=1$ can be explained as a consequence of the enormously-fluctuating random Coulomb potential created by randomly-positioned donor and acceptor impurities.  In later papers we extended this analysis to the case of near complete compensation \cite{Chen2013asr}, $K < 1$ and $1 - K \ll 1$, and we examined the effect of random Coulomb impurities on the surface disorder and transport properties \cite{Skinner2013trp}.  In this mini-review our goal is to outline in a general way the effects of random, 3D-distributed Coulomb impurities in TIs on both the bulk and surface properties.  We describe the screening mechanisms for the random Coulomb potential both within the bulk of the TI and at the surface, and we present predictions for the magnitude of the disorder potential and the conductivity. 

Our theoretical treatment is also motivated by the recent experiments of Ref.\ \onlinecite{Beidenkopf2011sfh}, where the random potential at the surface of typical TIs (Bi$_2$Se$_3$ and Bi$_2$Te$_3$) was studied directly by spectroscopic mapping with a scanning tunneling microscope.  It was shown that near the Dirac energy random fluctuations of the potential have a Gaussian-like distribution with a width $\sim 20$ -- $40$\,meV that can be attributed to deep impurity charges.  We show below that such fluctuations are consistent with disorder produced by three-dimensionally distributed bulk Coulomb impurities that are screened by the gapless TI surface.  

Crucial to our theoretical description throughout this paper is the assumption of a random spatial distribution of impurities.  This assumption is readily justified for TI samples made by cooling from a melt, where the distribution of impurities in space is a snapshot of the distribution that impurities have at higher temperature, when their diffusion practically freezes~\cite{Keldysh1964}. In 3D TIs, as in conventional narrow band gap semiconductors, at this temperature there is a concentration of intrinsic carriers larger than the concentration of impurities. Intrinsic carriers thus screen the Coulomb interaction between impurities, so that impurities remain randomly distributed in space. When the temperature is lowered to the point where intrinsic carriers recombine, the impurities are left in random positions~\cite{Galpern1972epc, Shklovskii1984epd}. 
If the diffusion of impurities freezes at $T \sim 1000$\,K it is reasonable to assume that impurities are randomly positioned for semiconductors with bulk band gap $E_g \leq 0.3$\,eV. Throughout this paper we deal with such narrow band gap TIs, such as Bi$_2$Se$_3$, for which our description of randomly-positioned impurities is accurate.  We also assume everywhere that donor and acceptor energy levels are shallow, meaning that their binding energy is much smaller than $E_g$.

The remainder of this paper can be divided into two parts.  In the first part, comprising Secs.\ \ref{sec:bulk} -- \ref{sec:Numerical}, we focus on bulk properties, essentially treating the TI as a strongly- or completely-compensated semiconductor and ignoring the surface states.  In Sec.\ \ref{sec:bulk} we give a conceptual explanation of the bulk disorder potential and the origin of the anomalously small bulk resistivity.  Sec.\ \ref{sec:model} formulates a numerical model of the TI bulk and uses it to calculate the corresponding electron density of states (DOS).  In Sec.\ \ref{sec:Numerical} we present our algorithm for computing the thermally activated conductivity, analyze our results, and arrive at an expression for the unusually small bulk activation energy.  We also evaluate the localization length of states close to the Fermi energy and estimate the characteristic temperature associated with variable-range hopping.

The second part of this paper, comprising Secs.\ \ref{sec:surfacetheory} -- \ref{sec:surfacegap}, deals with the effects of Coulomb impurities on the properties of the TI surface.  In Sec.\ \ref{sec:surfacetheory} we describe a self-consistent theory of the screened disorder potential at the TI surface and compare it with experiment. Sec.\ \ref{sec:surfaceconductivity} uses this theory to calculate the conductivity of surface electrons.  Sec.\ \ref{sec:surfacetobulk} briefly discusses how the amplitude of the disorder potential transitions from its large bulk value to its smaller value at the surface.  Finally, Sec.\ \ref{sec:surfacegap} discusses an extension of our analysis to the case where the TI surface has a gap introduced by some source of time-reversal symmetry breaking.  Where applicable, the major results of each section are summarized at the beginning of the section.

\section{Origin of the enhanced bulk conductivity}
\label{sec:bulk}

As mentioned in the Introduction, randomly-positioned impurities create a disordered Coulomb landscape in the bulk of the TI, which has the effect of reducing the activation energy $\Delta$ relative to what one would naively expect by thinking about flat valence and conduction bands.  In this section we explain this idea more fully, focusing first on the case of complete compensation, where the bulk transport can be described using the theory of a completely compensated semiconductor (CCS)~\cite{Shklovskii1972ccc, Shklovskii1984epd}. 

\begin{figure}[htb!]
\centering
\includegraphics[width=0.44 \textwidth]{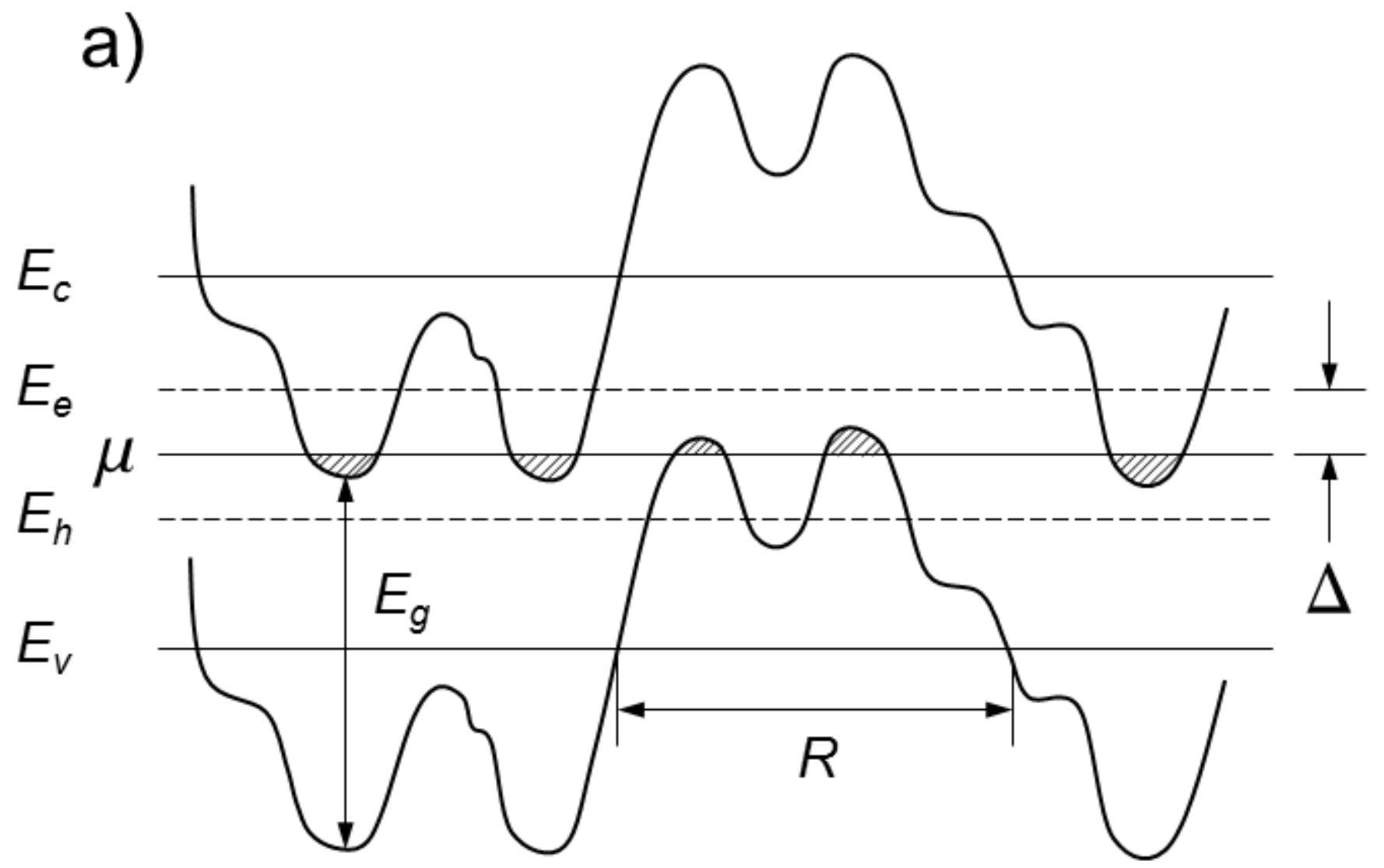}

\includegraphics[width=0.44 \textwidth]{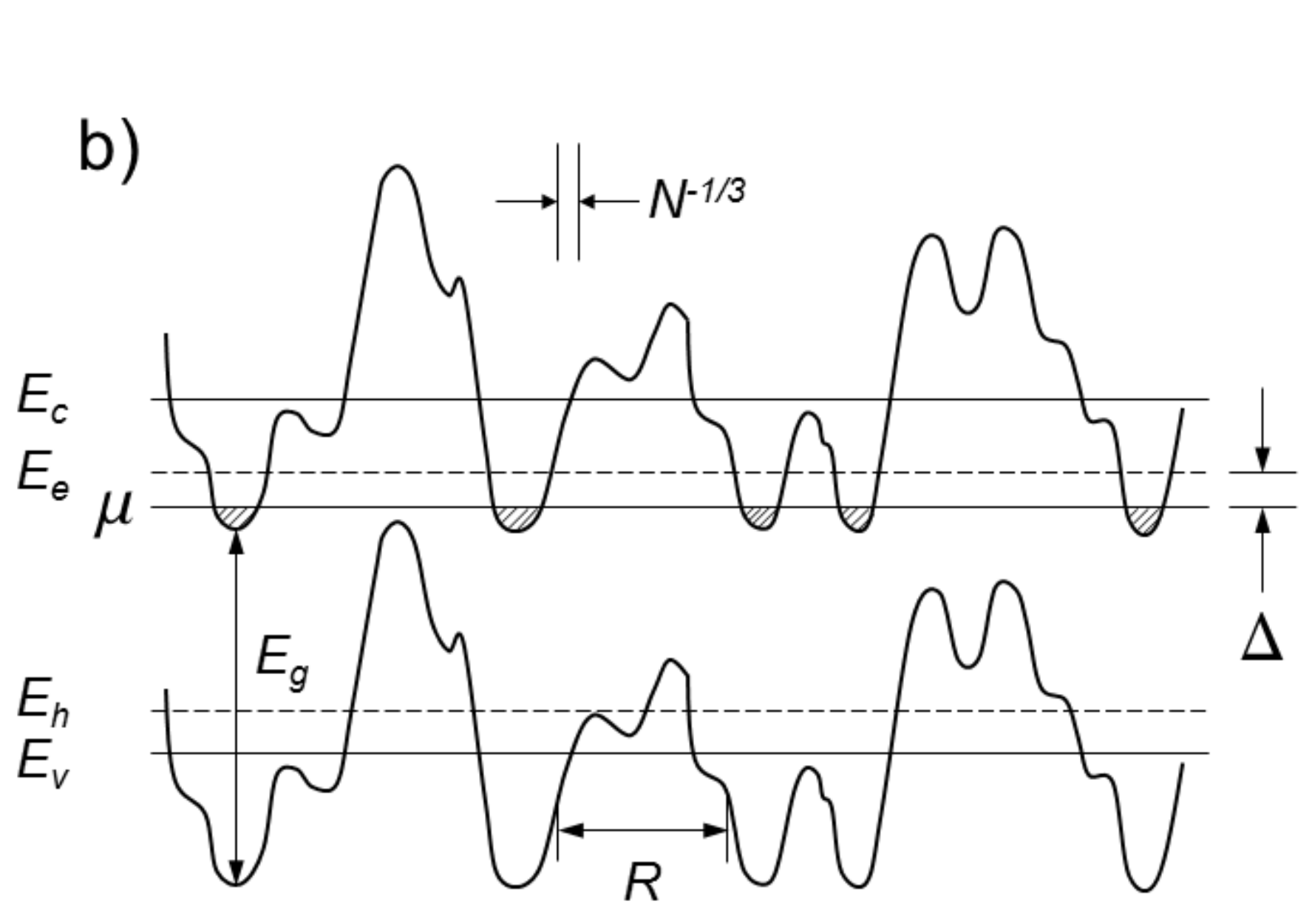}
\caption{Energy diagram of a) a completely compensated semiconductor ($K = 1$) and b) a strongly compensated semiconductor ($1-K \ll 1$) with band gap $E_g$. The upper and  the lower straight lines indicate the unperturbed positions of the bottom of the conduction band, $E_c$, and the ceiling of the valence band, $E_v$; the middle straight line corresponds to the Fermi level $\mu$. Meandering lines represent the band edges, which are modulated by the fluctuating potential of charged impurities.  $R$ is the characteristic size of potential fluctuations. Percolation levels (mobility edges) for electrons, $E_e$ and holes, $E_h$ are shown by dashed lines. Puddles occupied by carriers are shaded. Shallow impurities levels are not shown because they practically merge with band edges.}
\label{fig:band}
\end{figure}

This theory is based on the idea that at $K=1$, when almost all donors and acceptors are charged, random spatial fluctuations of the local concentration of impurities result in large fluctuations of charge. Their potential is poorly screened, because of the vanishing average concentration $n = N_D-N_A$ of electrons, and therefore has huge fluctuations. These fluctuations bend the conduction and valence band edges and in some places bring them to the Fermi level, creating electron and hole puddles that in turn non-linearly screen the random potential. As a result, the amplitude of potential fluctuations is limited by $E_g/2$, so that the ground state, illustrated schematically in Fig. \ref{fig:band}, resembles a network of $p$-$n$ junctions~\cite{Shklovskii1972ccc, Shklovskii1984epd}. The characteristic size of these $p$-$n$ junctions is \cite{Skinner2012wib}
\be
R = \frac {E_g^{2}\kappa^2}{8\pi Ne^{4}},
\label{eq:Rg}
\ee 
which can be thought of as the correlation length of the random potential.  For the typical parameters $E_g \sim 0.3$\,eV, $N_D = 10^{19}$ cm$^{-3}$, and dielectric constant $\kappa = 30$, this length scale  $R \approx 150$\,nm $\gg N_D^{-1/3} = 4.6$\,nm. That is, we deal with a very long range potential.  

As a result of these long range fluctuations, the resistivity can be dramatically different from the naive expectation based on thinking about flat valence and conduction bands. First, at relatively high temperatures activated conductivity is due to electrons and holes activated from the Fermi level to their corresponding classical percolation levels (classical mobility edges), $E_e$ and $E_h$, in the conduction and the valence bands. According to numerical modeling\cite{Skinner2012wib} at $K = 1$, the activation energy $\Delta \simeq 0.15 E_g$, meaning that $E_e$ and $E_h$ are substantially closer to the Fermi level $\mu$ than to the unperturbed bottom of the conduction band, $E_c$, or ceiling of the valence band, $E_v$ (Fig. \ref{fig:band}a).  ($E_c$ and $E_v$ are the energies of the conduction and valence bands, respectively, as they would be in the absence of a random potential.)  Thus, one can think of the universal small factor $\Delta/E_g \approx 0.15$ as corresponding to a percolation threshold associated with percolation through the potential created by random Coulomb impurities in 3D.

Second, at sufficiently low temperatures electrons and holes can hop (tunnel) directly between puddles, so that activated transport is replaced by VRH. In Ref.\ \onlinecite{Skinner2012wib} we showed that with decreasing temperature the activated resistivity crosses over directly to the Efros-Shklovskii (ES) law~\cite{Efros1975cga}
\be
\rho = \rho_{0}\exp(\Tes/T)^{1/2},
\label{eslaw}
\ee
where $\Tes = Ce^2/k_B\kappa\xi$, $e$ is the electron charge, $\xi$ is the localization length of electron states with energy close to the Fermi level, and $C \approx 4.4$ is a numerical coefficient. Together our results for the activated and VRH resistivity established the universal upper limit of the bulk resistivity $\rho(T)$ for a 3D TI compensated by shallow impurities.

In a Ref.\ \onlinecite{Chen2013asr}, we expanded our focus to consider not just the maximum possible bulk resistivity that appears at $K=1$, but to address the more practical question of the dependence of the bulk resistivity on the degree of compensation $K$ at $1-K \ll 1$. 
Indeed, with existing methods of growth of TI samples one cannot get  
$K=1$ exactly, and it is important to know how the results for a CCS, where $K=1$, are extended to the case of a strongly compensated semiconductor (SCS), for which $0 < 1-K \ll 1$. For example, one can ask at which value of $1-K$ does the activation energy $\Delta$ become twice smaller than at $K=1$. For definiteness we consider an $n$-type SCS, where the concentration of electrons $n = N_D - N_A \ll N_D$. We model numerically the ground state of such a SCS and its resistivity using algorithms similar to those of Ref.~\onlinecite{Skinner2012wib}.
We find that, in agreement with analytic theory~\cite{Shklovskii1984epd}, when $1-K$ grows the screening of the random potential improves and the correlation length $R$ of the random potential decreases.  The amplitude of the random potential decreases as well; hole puddles shrink and eventually vanish; and the chemical potential $\mu$ moves up, so that $E_c-\mu$ decreases. One can say that with increasing  $(1-K)$ screening happens by bending of the conduction band only, while all acceptors remain occupied by electrons and negatively charged. All these changes are illustrated by the transition from a) to b) in Fig. \ref{fig:band}. 

As a result of these changes with growing $1-K$, the activation energy $\Delta$ decreases. We find that the relation $\Delta = 0.3 (E_c-\mu) $ obtained in Ref.~\onlinecite{Skinner2012wib} for $K=1$  remains valid for $1-K \ll 1$ as well (see Fig. \ref{fig:EA} below). [In $p$-type semiconductors, where $K=N_D/N_A$, a similar relationship holds: $\Delta = 0.3 (\mu-E_v)$.] By $K = 0.97$ the activation energy $\Delta$ is already several times smaller than at $K=1$. This result shows that achieving the maximum bulk resistivity, with $\Delta = 0.15 E_g$, is not easy. It also helps to explain the origin of the large scatter in the magnitude of $\Delta$ among different TI samples~\cite{Ren2011o$s}.

Our prediction that $\Delta = 0.3 (E_c - \mu) $ can in principle be directly compared with experiments in TIs. Indeed, for each $K$ the position of the Fermi level, $(E_c-\mu)$, can be found via measurements of the concentration of electrons in the surface states using Shubnikov-de-Haas oscillations. 

At lower temperatures the activated bulk conduction crosses over to ES VRH.  In Sec.\ \ref{sec:Numerical} we study this crossover numerically and also show how $\Tes$, which is correlated with $\Delta$, decreases with $1-K$. 

It should be mentioned that these results for the bulk conductivity are also applicable to other narrow gap semiconductors, for example, to InSb. Historically, a large effort was made to make InSb insulating via strong compensation, with the goal of improving the performance of InSb-based photodetectors. Results were again frustrating: the dark resistivity was too small. Our results are in reasonable agreement with transport experimental data for InSb~\cite{Gershenzon1974cim, Yaremenko1975ltc}.

\section{Model of bulk impurities and the density of states}
\label{sec:model}

In order to study numerically the bulk properties of a heavily doped SCS, we introduce a model of the bulk donors and acceptors.  In this section we first describe our numerical model and then use it to calculate the position of the Fermi level relative to the band edges as a function of compensation, $K$, and to evaluate the density of states of impurity states.  Our major results are shown below in Figs.\ \ref{fig:MU} and \ref{fig:DOS}.

Specifically, we model the bulk as a cube containing a large number of randomly-positioned donors and acceptors.  We numerate all donors and acceptors by the index $i$ and use $n_i = 0$ or $1$ to denote the number of electrons residing on a donor or acceptor. We also introduce the binary variable $f_i$ to discriminate between donors (for which $f_i = 1$) and acceptors ($f_i = -1$). The Hamiltonian of our system is then
\be
H= \sum_i \frac{E_g}{2} f_i n_i +  \sum_{\langle ij\rangle} V(r_{ij}) q_i q_j,
\label{Hamiltonian}
\ee
where $q_i = (f_i/2 - n_i + 1/2)$ is the net charge of site $i$, $V(r)$ is the interaction energy between two like-charged impurities at a distance $r$, and all energies are defined relative to the Fermi level. The first term of Eq.\ (\ref{Hamiltonian}) contains the difference between the energies of donors and acceptors, which for the case of shallow impurities is very close to the semiconductor gap $E_g$.
The second term of $H$ represents the total interaction energy of charged impurities.  Note that Eq.\ (\ref{Hamiltonian}) does not include the kinetic energy of electrons and holes in the conduction and valence bands and, therefore, aims only at a description of the low temperature physics of SCS $(k_B T \ll E_c - \mu)$. 

The form of the interaction law $V(r)$ requires some consideration.  For two impurities at a distance $r \gg a_B$, where $a_B$ is the effective Bohr radius of impurity states, one can use for $V(r)$ the normal Coulomb interaction $V(r)=e^2/\kappa r$. For example, one can consider a pair of empty and distant donors.  In such a donor pair one donor shifts the energy of the electron level on the other by $V(r)= - e^2/\kappa r$. This classical form for $V(r)$ is good for a lightly doped SCS. In a heavily doped SCS, on the other hand, where $a_B  > N_{D}^{-1/3}$, most impurities have at least one neighbor at distance $r < a_B$, and quantum mechanical averaging over the electron wave function becomes important. (This is why an uncompensated heavily doped semiconductor is a good metal.) For example, a pair of donors cannot create an electron energy state deeper than that of the helium-like ion, which has binding energy $2e^2 /\kappa a_B$ is the binding energy of the shallow donor state.  The interaction law $V(r)$ should therefore be ``softened" at short distances $r < a_B$ to reflect quantum mechanical effects.  We model this behavior by continuing to use the classical Hamiltonian of Eq.\ (\ref{Hamiltonian}) with a truncated Coulomb potential $V(r)=e^2/\kappa (r^2 + a^2_B)^{1/2}$. 

Below it is convenient to express energies in units of $e^2 N_{D}^{1/3}/\kappa$.  In these units, a typical TI  with band gap $0.3$\,eV has $E_g \approx 30$.  We unfortunately could not model $E_g = 30$ directly, since in this case the very large correlation length of the random potential, $R$, leads to large size effects. Instead, we present results for the more modest value $E_g = 15$, for which the size effect requires extrapolation\cite{Skinner2012wib} only for $K=1$.  Results for the smaller $E_g = 10$ are largely identical \cite{Skinner2012wib}.
  
In our numerical simulations, we first randomly place donors and acceptors within the simulation volume; results presented below correspond to $M = 20000$ donors and $20000 K$ acceptors.
We then search for the arrangement of electrons (or equivalently, the set of electron occupation numbers $\{n_i\}$) that minimizes $H$, and we use this set to calculate the DOS and the conductivity.  We begin our search from the state where all $MK$ acceptors are populated by electrons and negative ($n_i=1, q_i=-1$), and where an equal number of randomly chosen donors are empty and positive ($n_i=0, q_i=1$), while the remaining $M(1-K)$ donors are filled and neutral ($n_i= 1, q_i= 0$).  The charged donors and acceptors in this initial state create a random potential whose magnitude exceeds $E_g$, and as a result the system's energy is well above that of the ground state. In order to bring the system closer to its ground state, we attempt sequentially to transfer electrons from an occupied impurity (either a neutral donor or a negatively charged acceptor) to an unoccupied one (a positively charged donor or a neutral acceptor). If the proposed move lowers the total system energy $H$, then it is accepted, otherwise it is rejected. To check whether $H$ goes down with each proposed move, for a given set of electron occupation numbers $\{n_i\}$ it is convenient to introduce the single-electron energy state, $\e_i$, at a given impurity $i$:
\be
\e_i = \frac{E_g}{2}f_i -  \sum_{j \neq i} V(r_{ij})q_j.  
\label{se}
\ee
In the ground state, single electron energies must satisfy the ES criterion
\be
\e_j - \e_i - V(r_{ij}) > 0
\label{eq:ES} 
\ee
for all $i$, $j$ with $n_i = 1$ and $n_j = 0$.  We use our numerical simulation to loop through all pairs of impurity sites $i$, $j$ and enforce this criterion; if a given pair does not satisfy Eq.\ (\ref{eq:ES}), then we move the electron from impurity $i$ to $j$ and recalculate all $\e_i$.  This process is continued until no single-electron transfers are possible that lower $H$. The final arrangement of electrons can be called a pseudo-ground state, since higher order stability criteria of the true ground state (corresponding to simultaneously changing three or more electron numbers) are not checked. Such pseudo-ground states are known to describe the properties of real ground states with a high degree of accuracy~\cite{Shklovskii1984epd, Mobius1992cgi}. Results below are obtained at $E_g=15$ and $a_B =N_D^{-1/3}$ for $K=1, 0.99, 0.98, 0.97, 0.96$ and $0.95$, and are averaged over 100 realizations of the impurity coordinates.

For each pseudo-ground state we estimate the Fermi energy $\mu$ as the arithmetic average of the minimum empty and maximum occupied energies $\e$.  The results are shown in Fig. \ref{fig:MU}, which shows how the Fermi level $\mu(K)$ shifts from the middle of the gap toward the conduction band bottom with growing $1-K$.
At  $1-K > 0.01$ this dependence is in reasonable agreement with the prediction of 
single band theory (which ignores the valence band and acceptors)~\cite{Shklovskii1984epd} that $E_c-\mu = A (1-K)^{-1/3}$, where $A$ is a numerical coefficient. Note, however that for heavily doped SCS the coefficient $A_h \simeq 1.4$ is twice smaller than the coefficient $A_l \simeq 2.8$ obtained in Ref.~\onlinecite{Shklovskii1984epd} for a lightly doped SCS, for which $N_D a_B^3 \ll 1$.  In the latter case the short range Coulomb interaction at distances $r \ll N_D^{-1/3}$ leads to an additional contribution to $\mu$ of the same order of magnitude.

\begin{figure}[tb!]
\centering
\includegraphics[width=0.42 \textwidth]{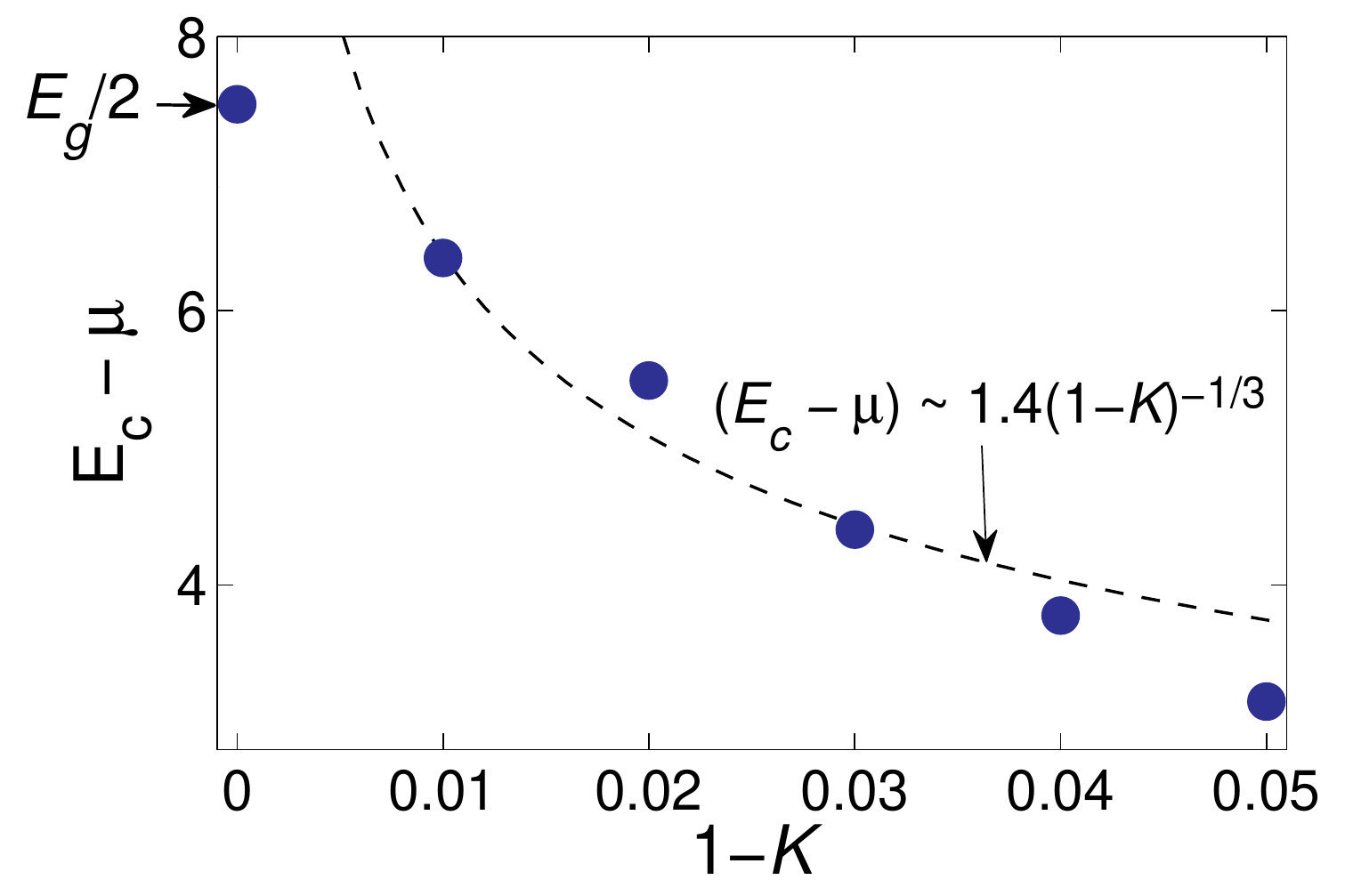}
\caption{(Color online) Distance between the Fermi level $\mu$ and the bottom of the conduction band $E_c$ as a function of $1-K$, as calculated by numerical simulation.  Energies are in units of $e^2 N_D^{-1/3}/\kappa$, and the simulated band gap is $E_g = 15$.  The size of dots characterizes the numerical uncertainty.}
\label{fig:MU}
\end{figure}

The resulting DOS of impurities is shown in Fig. \ref{fig:DOS} for $K = 1$ and $K=0.95$. $g^*(\e)$ is the DOS in the units of $(1+K)N_{D}/(e^2 N_{D}^{1/3}/\kappa)$ and is normalized to unity.  At $K = 1$, the nearly constant and symmetric DOS between $\e = -E_g$ and $\e = E_g$ reflects the practically uniform distribution of the random potential from $-E_g/2$ to $E_g/2$ and, correspondingly, of the band edges $E_c$ and $E_v$ between $0$ to $E_g$ and between $0$ to $-E_g$, respectively (see Fig. \ref{fig:band}a).  Near the Fermi level ($\e = 0$) one can see the ES Coulomb gap~\cite{Efros1975cga}. 

\begin{figure}[tb!]
\centering
\includegraphics[width=0.48 \textwidth]{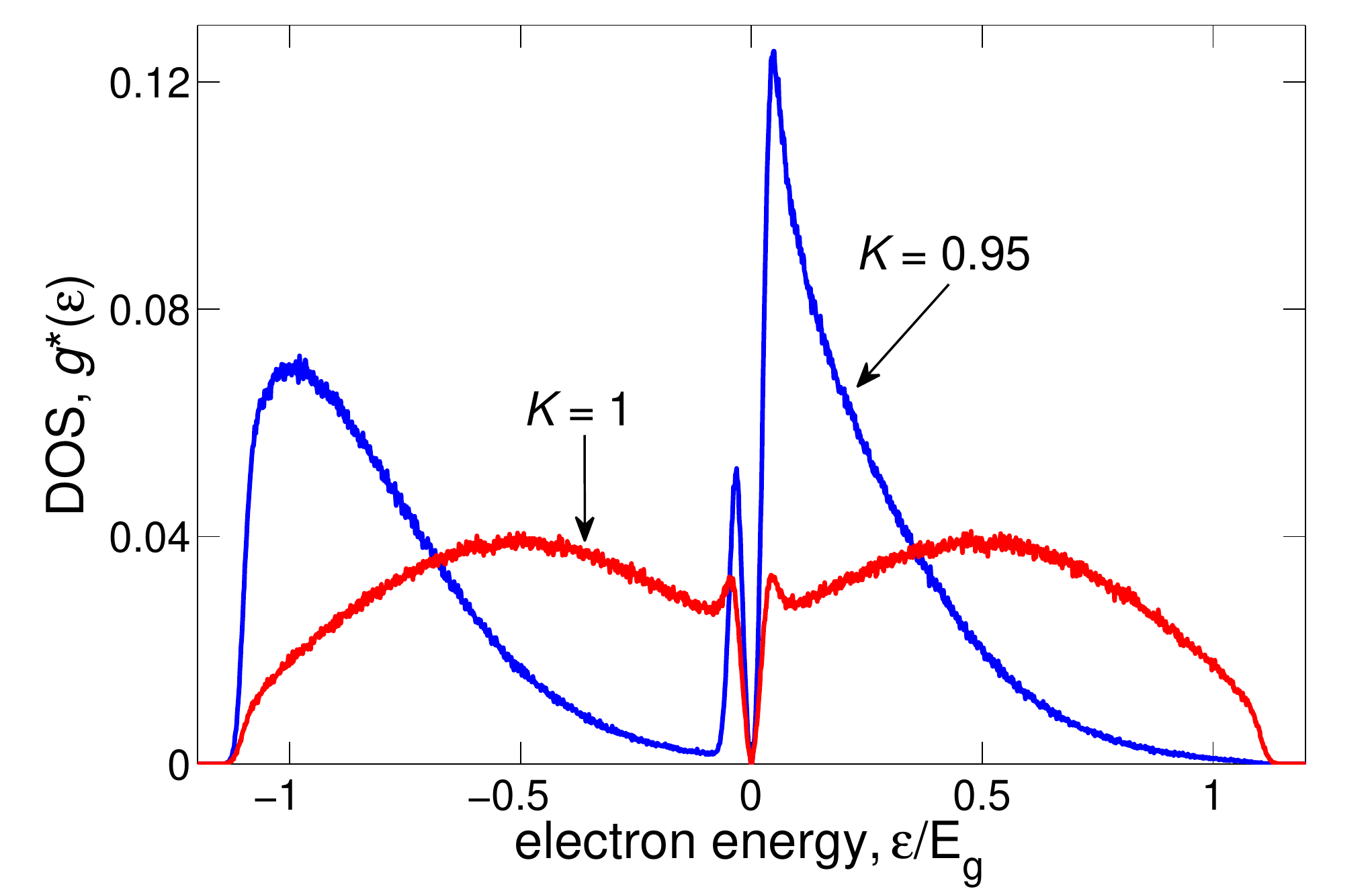}
\caption{(Color online) Dimensionless single-electron DOS $g(\e)$, in units of $[(1+K)N/(e^2N^{1/3}/\kappa)]$, as a function of electron energy $\e$ calculated from the Fermi level.  Results are plotted for $K$ = 0.95 (blue) and $K = 1$ (red) using $E_g = 15$. Impurity states with $\e <0$ are occupied and those with  $\e  > 0$ are empty.
At $K = 1$ the total DOS of impurities has donor-acceptor symmetry, which is lost with growing $1-K$.}
\label{fig:DOS}
\end{figure}

On the other hand, at $K < 1$ the DOS of impurity states loses the donor-acceptor symmetry it has at $K=1$.
As described in Sec.\ \ref{sec:bulk} (see Fig. \ref{fig:band}), with growing $1-K$ hole puddles are eliminated so that acceptors become disengaged from screening.  The acceptor DOS (leftmost peak of Fig.\ \ref{fig:DOS}) therefore splits from the donor one, which in turn develops two peaks separated by the Fermi level at $\e = 0$. The large right peak belongs to empty donors, while the small and narrow left peak belongs to occupied donors  (electron puddles).  These two donor peaks are separated by the ES Coulomb gap.

\section{Numerical modeling of thermally activated conductivity}
\label{sec:Numerical}

\begin{figure}[tb!]
\centering
\includegraphics[width=0.5 \textwidth]{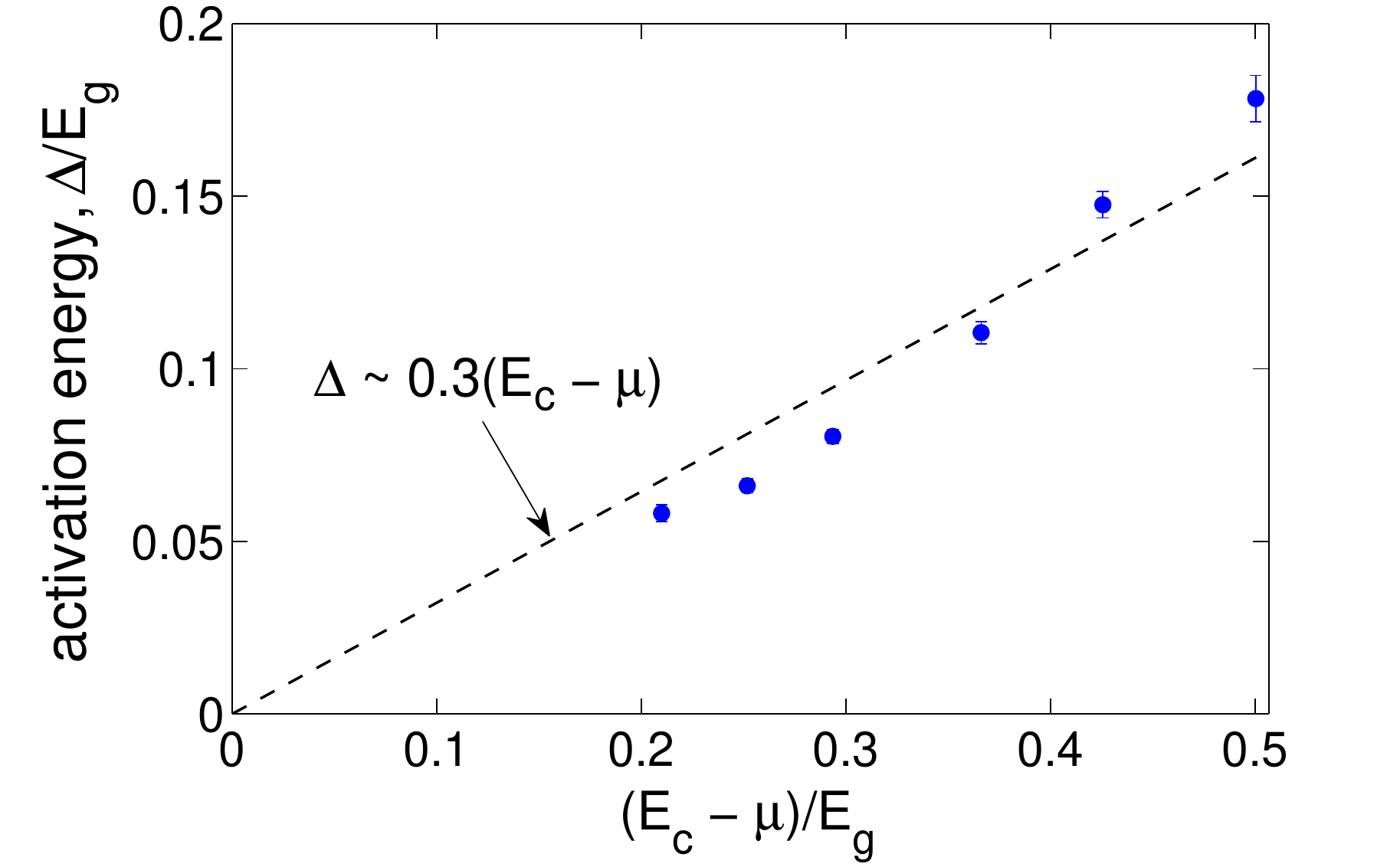}
\caption{(Color online) The activation energy $\Delta$ as a function of the distance between the Fermi level and the conduction band, plotted for $K = 1.0, 0.99, 0.98, 0.97, 0.96$, and $0.95$ (from right to left). The dashed line is the best linear fit, $\Delta \simeq 0.3(E_c - \mu)$.  All energies are plotted in units of the band gap $E_g$.}
\label{fig:EA}
\end{figure}

In the previous section we described our procedure for finding the energy levels of donor and acceptor impurities in the pseudo-ground state.  We now discuss how these results can be used to calculate the bulk conductivity of a SCS, and we present results for the conductivity both in the high-temperature, activated regime and in the low-temperature, VRH regime.  Our major results are twofold.  First, we find that in the activated regime the activation energy decreases as the chemical potential approaches the conduction band according to $\Delta \approx 0.3 (E_c - \mu)$ [see Fig.\ \ref{fig:EA}].  Second, we study how the characteristic temperature $\Tes$ in the VRH regime depends on compensation, and we find that $\Tes \simeq 4.4 \sqrt{\Delta (e^2 N_D^{1/3}/\kappa)}$.

Our process for numerically calculating the resistivity is as follows.  Once the energies $\{\e_i\}$ are known (as calculated using the procedure described in Sec.\ \ref{sec:model}), we evaluate the resistivity using the approach of the Miller-Abrahams resistor network~\cite{Miller1960ica, Shklovskii1984epd}.  In this description each pair of impurities $i, j$ is said to be connected by a link with resistance $R_{ij}=R_0\exp[2 r_{ij}/\xi + \e_{ij}/k_BT]$, where the activation energy $\e_{ij}$ is defined~\cite{Shklovskii1984epd} as follows: 
\be 
\e_{ij} = \left\{
\begin{array}{lr}
|\e_j - \e_i| - V(r_{ij}), &  \e_j\e_i < 0 \vspace{2mm} \\
\max \left[ \left|\e_i \right|, \left|\e_j \right| \right], &  \e_j\e_i > 0.
\end{array}
\right.
\label{eq:Eij}
\ee
The resistivity of the system as a whole is found using a percolation approach \cite{Shklovskii1984epd}.  Specifically, we find the minimum resistance $R_c$ such that if all links with resistance $R_{ij} > R_c$ are cut, then there still exists a percolation pathway connecting opposite faces of the simulation volume.  This approach captures the exponential dependence of the resistivity on the temperature, and we ignore details of the prefactor.  Below we plot the temperature in the dimensionless units $T^* = 2 k_B T \kappa / e^2 N_D^{2/3} \xi$ and the resistivity $\rho$ using the dimensionless quantity $(\ln\rho)^* = (\xi N_D^{1/3}/2) \ln R_c/R_0$.  These dimensionless units eliminate any explicit dependence on the localization length $\xi$.

\begin{figure}[tb!]
\centering
\includegraphics[width=0.5 \textwidth]{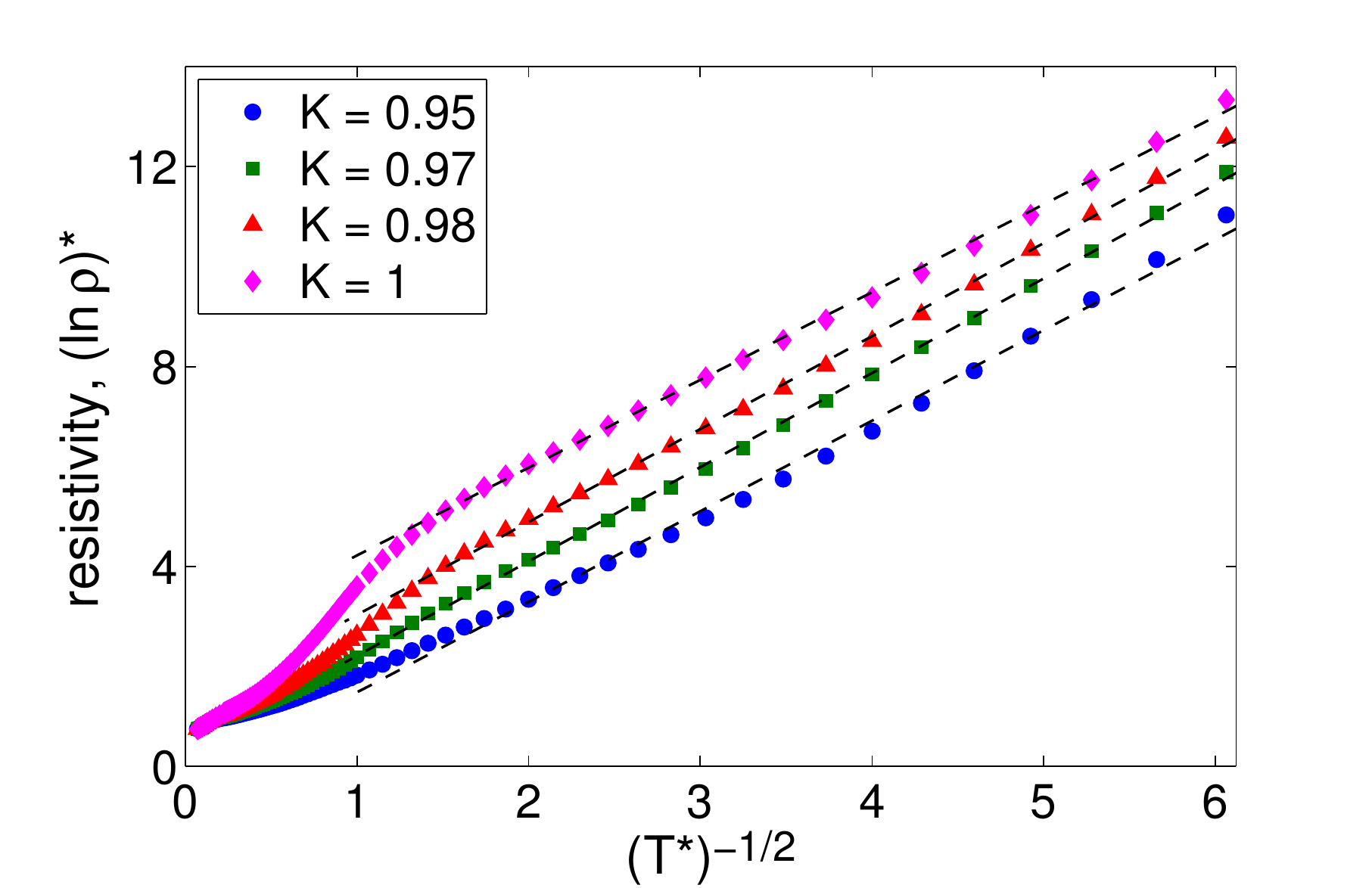}
\caption{(Color online) The temperature dependence of the resistivity in the whole temperature range $ 200 > T^* > 0.03$. The dimensionless resistance $(\ln \rho)^*$ is plotted against $(T^*)^{-1/2}$ to illustrate that the resistivity follows the ES law at low temperatures. The dashed lines are the best linear fits.}
\label{fig:RL}
\end{figure}

\begin{figure}[tb!]
\centering
\includegraphics[width=0.5 \textwidth]{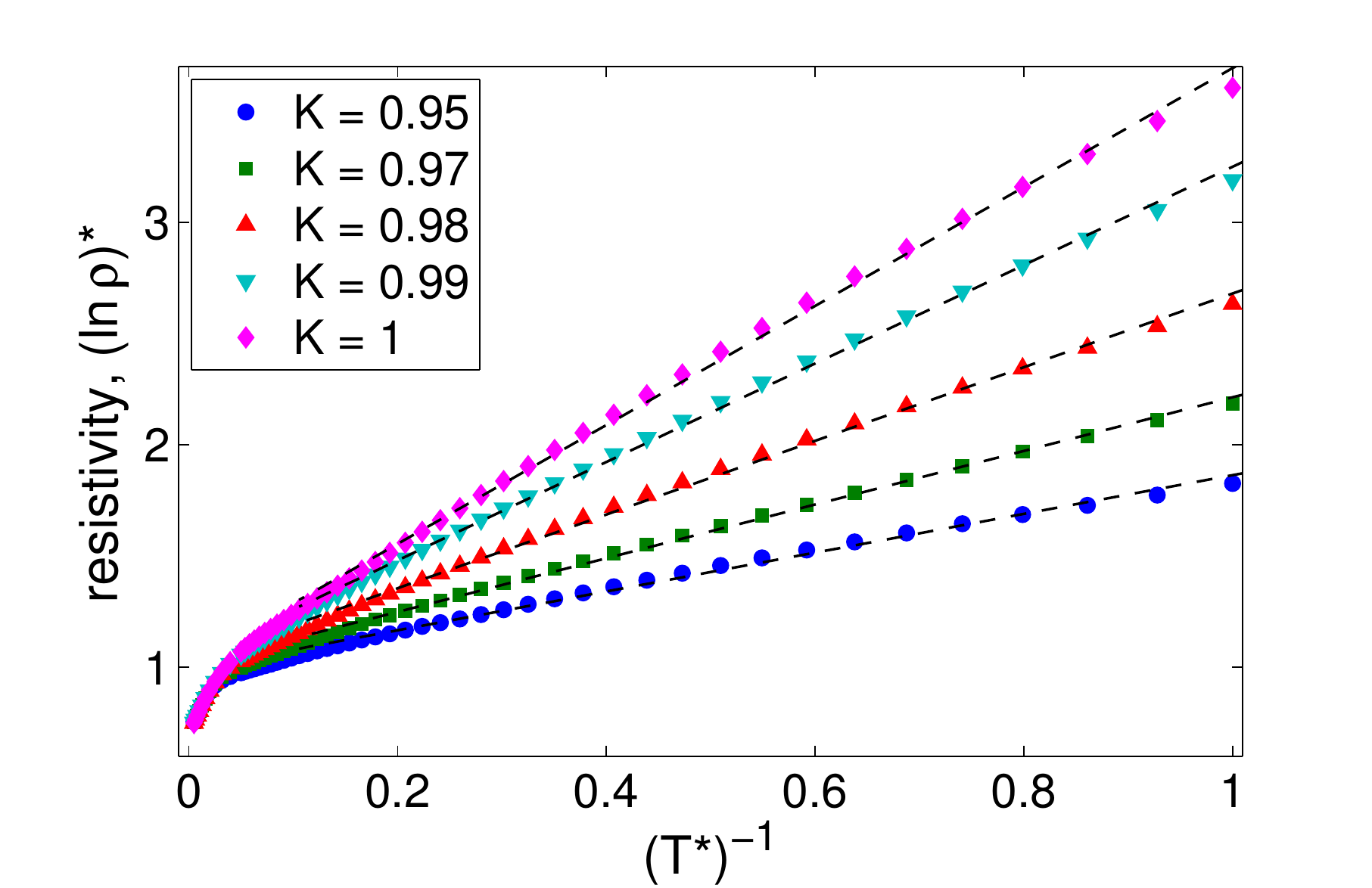}
\caption{(Color online) The temperature dependence of the resistivity in the high temperature range $ 200 > T^* > 1$. The dimensionless resistance $(\ln \rho)^*$ is plotted against $(T^*)^{-1}$ to illustrate that the resistivity is activated at high temperatures. The dashed lines are the best linear fits.}
\label{fig:RH}
\end{figure}

In Fig.\ \ref{fig:RL} the resulting resistivity is plotted as a function of $(T^*)^{-1/2}$ over the huge range of temperature $200 > T^* > 0.03$ for four different values of the compensation $K$.  The resulting linear dependence at $0.3 > T^* > 0.03$ indicates that at low temperatures the resistivity is well described by the ES law [see Eq.\ (\ref{eslaw})]. The higher temperature range $200 > T^* > 1 $ is plotted separately as a function of $1/T^*$ in Fig. \ref{fig:RH}. Here the linear slope suggests a well-defined activation energy that depends on the compensation $K$.  At extremely high $T^* \gtrsim 50$, which generally corresponds to unrealistically large temperatures, the conduction is dominated by activation of carriers across the band gap, which is not captured by our model.

Extracting the slope of the curves in Fig.\ \ref{fig:RH} (dashed lines) gives an estimate of the activation energy $\Delta$ as a function of compensation $K$.  Combining this result with the values for the chemical potential $\mu(K)$ calculated in Sec.\ \ref{sec:model} yields the data shown in Fig.\ \ref{fig:EA}, where $\Delta$ is plotted as a function of $(E_c-\mu)$ for all the studied values of compensation $K = 1, 0.99, 0.98, 0.97, 0.96, 0.95$.  
One can see that the equation $\Delta \simeq 0.3(E_c-\mu)$ holds reasonably well for all $K$ in this interval.

So far we have emphasized results that do not explicitly depend on the localization length $\xi$. In fact, $\xi$ determines the magnitude of $\Tes$, and therefore determines the value of temperature at which the conduction transitions from activated to VRH behavior. We argue now that in a TI $\xi$ is quite large, leading to a prominent role for VRH.  To see this, one can imagine an electron with energy close to the Fermi level tunneling from one electron puddle to another, distant one.  If such an electron were to tunnel along the straight line connecting the two puddles it would tunnel through high barriers and its wave function would decay sharply, with a decay length $\xi \ll a_B$.  However, this straight line does not constitute the path of least action for the tunneling electron.  Instead, a tunneling electron can use the same geometrical path as a classical percolating electron, which has energy $\Delta$ above the Fermi level, and thereby avoid large barriers. One can roughly estimate the tunneling decay length by assuming that along such a ``percolating" tunneling path the potential energy barriers $V$ are uniformly distributed in the range $0 \leq V \leq \Delta$ and neglecting the additional contribution to the action associated with curvature of this path. Integration over $V$ then gives a localization length $\xi \sim \hbar /\sqrt{m \Delta}$ and $k_B \Tes = 4.4 (m \Delta)^{1/2} (e^2/\kappa \hbar) $.  For a TI with $a_B = N_D^{-1/3}$ this implies $k_B \Tes = 4.4 \sqrt{ \Delta (e^{2} N_D^{1/3}/\kappa) }$. 

The dependence $\Tes \propto \sqrt{\Delta}$ implies that when $\Delta$ increases $\sim 2.5$ times, as in Fig.\ \ref{fig:EA} corresponding to the difference between $K = 0.95$ and $K = 1$, the ES temperature $\Tes$ increases by $\sim 60\%$. For a TI with $\kappa = 30$ and $N_D = 10^{19}$ cm$^{-3}$, this corresponds to a variation in $\Tes$ from 500 to 800\,K.  The regime of ES VRH in TIs can be studied experimentally, but such a study requires sufficiently thick samples that the bulk conduction provides a larger contribution to the total conductance than the TI surfaces.

\section{Self-consistent theory of the surface disorder potential}
\label{sec:surfacetheory}

In the first part of this paper we showed how the bulk conduction is strongly influenced by the presence of random Coulomb impurities, which produce large bending of the bulk conduction and valence bands.  We now turn our attention to the problem of how these same impurities affect the surface transport provided by the Dirac-like surface states.  For this problem we adopt the same model of monovalent Coulomb impurities that are randomly distributed throughout the bulk of the TI, and we focus our attention on the case of complete (or nearly-complete) compensation $N_D = N_A \equiv N$, where the Fermi level lies within the bulk band gap.  As we show below, for determining the properties of the surface one can safely ignore the weak nonlinear screening by electron and hole puddles formed in the bulk (illustrated in Fig.\ \ref{fig:band}).

In this section we present a self-consistent theory for the magnitude of the disorder potential at the TI surface, following Ref.\ \onlinecite{Skinner2013trp}.  Our primary result is an expression for the amplitude of fluctuations of the electric potential energy, $\Gamma$, at the TI surface as a function of the chemical potential, $\mu$, measured relative to the Dirac point.  In particular, for $\mu = 0$ we show below that 
\be 
\Gamma^2 = \frac{ \sqrt[3]{2} \pi}{\alpha^{4/3}}  \left( \frac{e^2 N^{1/3}}{\kappa_s} \right)^2,  \hspace{5mm} (\mu = 0).
\label{eq:gamma0}
\ee 
Here $\alpha = e^2/\kappa_s \hbar v$ is the effective fine structure constant, where $\kappa_s$ is the effective dielectric constant at the surface and $v$ is the Dirac velocity.  This expression describes screening of the disorder potential via the formation of electron and hole puddles at the TI surface.  The characteristic size of these puddles is given by
\be 
r_s = \frac{N^{-1/3}}{2^{2/3} \alpha^{4/3}},  \hspace{5mm} (\mu = 0),
\label{eq:rs0units}
\ee
and the corresponding total number of electrons (or holes) per unit area in surface puddles is given by
\be 
n_p = \left( \frac{\alpha}{16} \right)^{2/3} N^{2/3},  \hspace{5mm} (\mu = 0).
\label{eq:ne}
\ee
Eqs.\ (\ref{eq:gamma0}) -- (\ref{eq:ne}) are derived below, along with results corresponding to large $\mu$.
Below we also derive a simple relation for the autocorrelation function of the potential at the TI surface, which has an unusually slow decay and can be used to verify the bulk origin of disorder.
These results were confirmed by numerical simulation in Ref.\ \onlinecite{Skinner2013trp}.

Our primary tool for describing screening of the electric potential is the Thomas-Fermi (TF) approximation, which applies in the limit where the potential $\phi(\rr)$ varies slowly compared to the characteristic Fermi wavelength of electrons at the surface.  Specifically, the TF approximation gives
\be 
\mu = E_f[n(\rr)] - e \phi(\rr),
\label{eq:TF}
\ee 
where $E_f(n) = \hbar v \sqrt{4 \pi |n|}\sgn(n) = (e^2/\alpha \kappa_s)\sqrt{4\pi |n|} \sgn(n)$ is the local Fermi energy and $n(\rr)$ is the 2D electron concentration at the point $\rr$ on the surface.  The TF approximation is justified whenever $\alpha \ll 1$, as we show below.  In TIs such small $\alpha$ can be seen as the result of the large bulk dielectric constant $\kappa \gtrsim 30$.  We note here that for describing the properties of the surface state, which exists  at a dielectric discontinuity, one should use for the effective dielectric constant $\kappa_s$ the arithmetic mean of the internal and external dielectric constants.  If the TI is in vacuum, then $\kappa_s = (\kappa + 1)/2 \simeq \kappa/2$.  

When the chemical potential is large enough in magnitude that $\mu^2 \gg e^2 \langle \phi^2 \rangle$, where $\langle ... \rangle$ denotes averaging over the TI surface, the relation $E_f(n)$ can be linearized to read $E_f[n(\rr)] \simeq \mu + \delta n(\rr)/\nu(\mu)$.  Here $\delta n(\rr) = n(\rr) - n_0$ is the difference in the electron concentration relative to the state with zero electric potential, $n_0 = \alpha^2 \kappa_s^2 \mu^2/(4 \pi e^4)$, and $\nu(\mu) = \alpha^2 \kappa_s^2 |\mu|/(2 \pi e^4)$ is the density of states at $E_f = \mu$.  From this density of states one can define a screening radius $r_s = \kappa_s/2\pi e^2 \nu = e^2/\alpha^2 \kappa_s \mu$ that characterizes the distance over which fluctuations in the Coulomb potential are screened by the surface.  The TF approximation is valid when the Fermi wavelength $\lambda_f \sim n_0^{-1/2} \sim e^2/\alpha \kappa_s \mu$ is much smaller than $r_s$, which gives the condition $\alpha \ll 1$.

One can understand qualitatively the magnitude of the potential fluctuations, $\Gamma$, using the following simple argument.  For a given point on the TI surface, one can say that only impurities within a distance $R' \lesssim r_s$ contribute to the potential; those impurities at a distance $R' \gg r_s$ are effectively screened out (one can say that they are screened by their image charges in the ``metallic" TI surface).  Impurities with $R' < r_s$, on the other hand, are essentially unscreened.  There are $\sim N r_s^3$ such impurities, and their net charge is of order $Q \sim e \sqrt{N r_s^3}$, with a random sign.  The absolute value of the potential at the surface is then $\sim Q/\kappa_s r_s$, so that $\Gamma \sim e Q/\kappa_s r_s \sim (e^2 N^{1/3}/\kappa_s) (N r_s^3)^{1/6} \sim \sqrt{e^2 N/\kappa_s \nu} \sim \sqrt{e^4 N/\alpha^2 \kappa_s^3 |\mu|}$.

In order to more accurately derive the value of $\Gamma$, one can start by considering the potential created by a single impurity charge $+e$.  When such an impurity charge is placed a distance $z$ from the TI surface (say, above the origin), it creates a potential $\phi_1(r; z)$ that within the TF approximation is given by \cite{Ando1982epo}
\be 
\phi_1(r; z) = \frac{e}{\kappa_s} \int_0^\infty \frac{\exp [-q z]}{1 + (q r_s)^{-1}} J_0(q r) \, dq ,
\label{eq:p1}
\ee 
where $J_0(x)$ is the zeroth order Bessel function of the first kind.  At large $z/r_s$, Eq.\ (\ref{eq:p1}) can be expanded to give 
\be 
\phi_1(r; z) \simeq \frac{e}{\kappa_s} \frac{z r_s}{(r^2 + z^2)^{3/2}}.
\label{eq:p1large}
\ee
A simple physical derivation of Eq.\ (\ref{eq:p1large}) is based on the notion \cite{Loth2009nsb} that for a distant impurity, such that $z \gg r_s$, a surface with screening radius $r_s$ effectively plays the role of a metallic surface positioned below the real surface at a distance $z = -r_s/2$.  Equation (\ref{eq:p1large}) can then be viewed as the sum of the potentials created by the original charge at a distance $z$ above the plane and its opposite image charge at a distance $z + r_s$ below the plane, expanded to lowest order in $r_s/z$.

The total potential at the origin is $\phi(0) = \sum_i q_i \phi_1(r_i; z_i)$, where the index $i$ labels all impurity charges, $q_i$ is the sign of impurity $i$, and $\rr_i$ and $z_i$ are the radial and azimuthal coordinates of its position.  Under the assumption that all impurity positions are uncorrelated and randomly-distributed throughout the bulk of the TI, the average of $\phi^2$ is given by
\be 
\langle \phi^2 \rangle = \int [\phi_1(r'; z')]^2 \, 2 N d^2 \rr' dz'.
\label{eq:gammaint}
\ee 
Here, the quantity $2N d^2\rr' dz'$ describes the probability that the volume element $d^2\rr' dz'$ contains an impurity charge, and the integration is taken over the semi-infinite volume of the bulk of the TI.  The width of the disorder potential at the TI surface, $\Gamma$, is defined by $\Gamma^2 = e^2 \langle \phi^2 \rangle$.  Inserting Eq.\ (\ref{eq:p1}) into Eq.\ (\ref{eq:gammaint}) and taking the integral then gives
\be 
\Gamma^2 = \frac{e^2 N}{\kappa_s \nu} = \frac{2 \pi e^4 N}{\alpha^2 \kappa_s^3 |\mu|}, \hspace{5mm} \left( |\mu| \gg \frac{e^2 N^{1/3}}{\kappa_s \alpha^{2/3}} \right). \label{eq:gammalargemu}
\ee 
Eq.\ (\ref{eq:gammalargemu}) is correct so long as the fluctuations in the Coulomb potential energy are small compared to the chemical potential, or $\Gamma \ll |\mu|$; this gives the condition written in parentheses.

On the other hand, when $|\mu|$ is very small, the fluctuations in the Coulomb potential become large compared to the chemical potential, and one cannot talk about a spatially uniform local density of states $\nu$ or screening radius $r_s$.  Instead, the Fermi energy has strong spatial variations, and the random potential is screened by the formation of electron and hole puddles at the surface.  Nonetheless, one can define an average density of states $\tdos$ at the surface, which determines, self-consistently, the typical screening radius $r_s$ and the magnitude of the potential fluctuations at the TI surface.  This value $\tdos$ can be equated with the \emph{thermodynamic} density of states of the system, $d \mu/d\langle n \rangle$, where $\langle n \rangle$ is the overall electron concentration of the surface.

Consider, for example, the case $\mu = 0$, where by symmetry the average value of the potential $\langle \phi \rangle = 0$.  At any given point $\rr$ on the surface, the potential $\phi(\rr)$ is the sum of contributions from many individual impurity charges, provided that the characteristic screening radius $r_s = \kappa_s/2\pi e^2 \tdos \gg N^{-1/3}$.  This implies that, by the central limit theorem, the value of the potential across the surface is  Gaussian-distributed with some variance $\langle \phi^2 \rangle = \Gamma^2/e^2$ that remains to be calculated.  Within the TF approximation the local density of states at the point $\rr$ is $\nu[-e\phi(\rr)] = e \alpha^2 \kappa_s^2 |\phi(\rr)|/(2 \pi e^4)$, so that one can calculate the average density of states as
\begin{eqnarray} 
\tdos & = & \int_{-\infty}^{\infty} \nu(-e\phi) \frac{\exp\left[ -e^2\phi^2/2 \Gamma^2 \right]}{\sqrt{2 \pi \Gamma^2/e^2}}  \, d\phi \nonumber \\
& = & \frac{\alpha^2\kappa_s^2 \Gamma}{\sqrt{2 \pi^3} e^4},  \hspace{5mm} (\mu = 0).
\label{eq:nu0}
\end{eqnarray}
This result for $\tdos$ can be inserted into the first equality of Eq.\ (\ref{eq:gammalargemu}), $\Gamma^2 = e^2 N/\kappa_s \tdos$, to give a self-consistent relation for the amplitude of potential fluctuations \cite{Stern1974ltl}.  This procedure gives the result first announced at the beginning of this section, Eq.\ (\ref{eq:gamma0}).
Substituting Eqs.\ (\ref{eq:gamma0}) and (\ref{eq:nu0}) into the expression for the screening radius, $r_s = \kappa_s/2\pi e^2 \tdos$, gives Eq.\ (\ref{eq:rs0units}).

One can also calculate the total concentration of electrons/holes in surface puddles, $n_p$, implied by this result for $\Gamma^2$.  This is done by first inverting the TF relation, Eq.\ (\ref{eq:TF}), at $\mu = 0$ to give $n(\phi) = (\alpha^2 \kappa_s^2/4 \pi e^2) \phi^2 \sgn(\phi)$.  Integrating this expression for $n(\phi)$ weighted by the Gaussian probability distribution for $\phi$ gives
\begin{eqnarray} 
n_p & = & \int_0^{\infty} n(\phi) \frac{\exp\left[ -e^2\phi^2/2 \Gamma^2 \right]}{\sqrt{2 \pi \Gamma^2/e^2}} d \phi \nonumber \\
& = & \frac{\alpha^2 \kappa_s^2 \Gamma^2}{8 \pi e^4},  \hspace{5mm} (\mu = 0). \nonumber
\nonumber 
\end{eqnarray}
Substituting the result of Eq.\ (\ref{eq:gamma0}) for $\Gamma^2$ then gives Eq.\ (\ref{eq:ne}).  One can also combine this result for the residual electron/hole concentration, $n_p$, with the expression for the screening radius, $r_s$, to arrive at an estimate for the number of electrons/holes per puddle: $M_p \sim \pi n_p r_s^2 \sim \pi/16 \alpha^2$.  Apparently at small $\alpha$ puddles typically contain many electrons/holes, $M_p \gg 1$.

Our primarily results, outlined in Eqs.\ (\ref{eq:gamma0}) -- (\ref{eq:ne}), are valid within the TF approximation so long as the typical Fermi wavelength, $\lambda_f \sim e^2/\alpha \kappa_s \Gamma$, is much smaller than the typical screening radius, $r_s \sim  e^2/\alpha^2 \kappa_s \Gamma$, which again gives the condition $\alpha \ll 1$. 

As we mentioned above, at $\mu = 0$ the screening radius $r_s$ describes the characteristic size of electron or hole puddles at the TI surface.  More generally, $r_s$ plays the role of a length scale over which potential fluctuations at the surface are correlated.  Such correlations can be discussed in a quantitative way by defining the potential auto-correlation function:
\be 
C(r) = \langle \phi(\RR') \phi(\rr' + \rr) \rangle_{\rr'} ,
\label{eq:Cdef}
\ee
where $\langle ... \rangle_{\rr'}$ denotes averaging over the spatial coordinate $\rr'$, and where by symmetry the correlation function depends on $|\rr| = r$ only.  In the remainder of this section we derive approximate analytical results for $C(r)$, and show that spatial correlations in the potential have an unusually slow decay.

At $r = 0$, Eq.\ (\ref{eq:Cdef}) reproduces the definition of $\langle \phi^2 \rangle$, so that $C(0) = \Gamma^2/e^2$.  At small enough distances that $r \ll r_s$, one can expect that the value of $C(r)$ is determined primarily by unscreened impurities that are within a distance $r_s$ from the surface, as explained above during the derivation of $\Gamma^2$.  On the other hand, at $r \gg r_s$ correlations are produced primarily by impurities that are relatively far from the surface, as can be seen from the following scaling argument.  Consider two surface points separated by a distance $r \gg r_s$.  One can imagine drawing a cube of size $r$ that extends into the bulk of the TI and which contains the two surface points on opposite edges of one of its faces.  Such a cube contains $\sim N r^3$ impurities, and has a net impurity charge with magnitude $q \sim e \sqrt{N r^3}$ and random sign.  These impurity charges are located at a mean distance $\sim r \gg r_s$ above the surface and, therefore, by Eq.\ (\ref{eq:p1large}), contribute a net potential $\sim q r_s /\kappa_s r^2 \sim (e/\kappa_s) \sqrt{N r_s^2/r}$ to both surface points.  The square of this potential roughly gives the autocorrelation of the potential, $C(r) \sim e^2 Nr_s^2 /\kappa_s^2 r$.

A more careful expression for $C(r)$ can be derived by writing
\be 
C(r) = \int \phi_1(\rr'; z') \phi_1(\rr'-\rr; z') 2N d^2\rr' dz',
\label{eq:Cint}
\ee 
similar to Eq.\ (\ref{eq:gammaint}).  Inserting the asymptotic expression of Eq.\ (\ref{eq:p1large}) for $\phi_1$ and evaluating the integral gives
\be 
C(r) \simeq \frac{2 \pi e^2 N r_s^2}{\kappa_s^2 r} = \frac{\Gamma^2/e^2}{r/r_s}, \hspace{5mm} (r/r_s \gg 1).
\label{eq:C}
\ee
This result was also confirmed by numerical simulation in Ref.\ \onlinecite{Skinner2013trp}.

Eq.\ (\ref{eq:C}) implies an unusually slow decay of potential correlations at the surface, which, as explained above, arises from long-range fluctuations of the potential created by deep bulk impurities.  This behavior can be contrasted with the much faster decay of $C(r)$ that would result from a two-dimensional (2D) distribution of Coulomb impurities at a distance $d$ from the surface:  
\footnote{This result can be obtained by replacing the bulk impurity charge density $2N$ in Eq.\ (\ref{eq:Cint}) with $n_i \delta(z-d)$.}
$C(r) \sim e^2 n_i d r_s^2 / \kappa_s^2 r^3$, where $n_i$ is the 2D impurity concentration.  Thus, by studying $C(r)$ experimentally by scanning tunneling microscopy, one can discriminate between disorder by bulk impurities and disorder by impurities located in a layer close to the surface.

We now discuss the magnitude of $\Gamma$ and $r_s$ implied by these expressions for typical TIs, which generally have an impurity concentration $N \sim 10^{19}$ cm$^{-3}$.  Typical values of the Dirac velocity and fine structure constant for TIs can be taken from Ref.\ \onlinecite{Beidenkopf2011sfh}, which reports $\hbar v =  1.3$ eV\,\AA\, and estimates $\alpha = 0.24$.  Using these parameters gives $\Gamma \sim 30$\,meV and $r_s \sim 20$\,nm at the Dirac point, $\mu = 0$.  At large $|\mu| \gtrsim 30$\,meV, both $\Gamma^2$ and $r_s$ decay as $1/|\mu|$.  

As mentioned in the Introduction, the theory presented in this section provides a good description of the recent experimental results of Ref.\ \onlinecite{Beidenkopf2011sfh}, where the random potential at the surface of the 3D TIs Bi$_2$Se$_3$ and Bi$_2$Te$_3$ was studied using a scanning tunneling microscope \cite{Skinner2013trp}.  Indeed, in these experiments it was found that the electric potential at the surface was well-characterized by a Gaussian distribution with a standard deviation $\Gamma \sim 10$ -- $20$\,meV, and the characteristic length scale of potential fluctuations was estimated as $r_s \sim 20$ -- $30$\,nm.  One can compare these measurements to our theoretical predictions by using the parameters listed above and inserting the measured chemical potential $\mu \sim 100$\,meV into Eq.\ (\ref{eq:gammalargemu}).  This procedure gives $\Gamma \sim 18$\,meV, and the corresponding screening radius $r_s \sim 5$\,nm, so that our theory is indeed in reasonably good agreement with experiment.  Further, Ref.\ \onlinecite{Beidenkopf2011sfh} found that the disorder potential at the surface was not correlated with the position of surface impurities, indicating that the surface disorder potential originates primarily from impurities deep below the TI surface, as we have described.

Throughout this section, we have worked within the assumption that bulk impurities are completely ionized, or in other words that there is no screening by conduction band electrons or valence band holes in the bulk.  Such an assumption is valid when the chemical potential resides in the middle of a large bulk band gap.  In this case donors or acceptors can only be neutralized by very large band bending discussed in Sec.\ \ref{sec:bulk} (see Fig.\ \ref{fig:band}).  Such fluctuations take place over a long length scale $R$ that scales as the square of the distance between the Fermi level and the nearest band edge [see Eq.\ (\ref{eq:Rg})] and is typically on the order of hundreds of nanometers for typical TIs \cite{Skinner2012wib}.  On the other hand, near the surface of the TI the potential fluctuations are screened much more effectively and over a much shorter distance, $r_s$, by the (ungapped) surface states.  As shown above, $r_s$ is typically $\lesssim 20$\,nm, and the amplitude of surface potential fluctuations $\Gamma \sim 30$\,meV $\ll E_g \sim 300$\,meV.  One can therefore safely assume that near the surface there is no large band bending and one can indeed treat bulk impurities as completely ionized.  The effect of bulk screening should appear only in the long-range behavior of the correlation function, $r \gg R$, where the $1/r$ decay of $C(r)$ is truncated and, as one can show, is replaced with $C(r) \sim e^2 N R r_s^2/\kappa_s^2 r^2$.

Finally, we note that our theory ignores the possibility of screening by material \emph{outside} the TI.  For example, if the TI is placed next to a metal electrode or an ionic liquid \cite{Xiong2012tqo}, then this external material can screen the large potential fluctuations created by the bulk, thereby decreasing $\Gamma$ and $r_s$.

\section{From surface to bulk}
\label{sec:surfacetobulk}

In Sec.\ \ref{sec:bulk} we showed that deep within the bulk of the TI the disorder potential has large fluctuations of order $\Gamma \sim E_g$.  On the other hand, in Sec.\ \ref{sec:surfacetheory} we showed that at the TI surface the disorder potential has a much smaller amplitude, $\Gamma \sim (e^2 N^{1/3}/ \kappa \alpha^{2/3})$.  
In this section we elaborate briefly on the crossover between these two results, or in other words we describe how the amplitude of potential fluctuations grow as one moves from the surface of the TI into the bulk.  

Generally speaking, as one moves a distance $z > 0$ into the bulk of the TI, the amplitude of the disorder potential increases in magnitude.  In order to see quantitatively how $\Gamma$ grows as a function of $z$, one can assume, for the moment, that the TI surface is equivalent to a perfect metallic plane.  In this case, each impurity at position $(\rr', z')$ has a corresponding image charge at $(\rr', -z')$, and the total potential at $(0, z)$ is equal to the sum of the potentials created by the original impurity and its image.  One can calculate $\Gamma^2(z)$ by averaging the square of this potential over all possible positions of the impurity charge [as in Eq.\ (\ref{eq:gammaint})].  This calculation gives $\Gamma^2(z) = 8 \pi N e^4 z/\kappa^2$.  That is, $\Gamma^2(z)$ grows linearly with the distance $z$ from the TI surface.  This growth continues until $\Gamma$ becomes large enough that $\Gamma^2(z) = (E_g/2)^2$, at which point electron and hole puddles begin to form in the bulk and one arrives at the bulk screening picture described in Ref.\ \onlinecite{Skinner2012wib}.  This distance corresponds to $z = R/4$; at smaller $z$ the potential fluctuations are small enough that practically all donors and acceptors are charged.  

One can now recall that the TI surface is not perfectly metallic, and that its screening length $r_s$ is finite, so that $\Gamma^2(z)$ should be somewhat larger.  In fact, at $z \gg r_s$ one can still use the formula above for $\Gamma^2(z)$ by introducing a small modification allowing for the fact that the metallic surface is effectively shifted to the position $z = -r_s/2$ (as discussed in Sec.\ \ref{sec:surfacetheory}).  Making this adjustment gives $\Gamma^2(z) = 8 \pi N e^4 (z + r_s/2)/\kappa^2$ at $z \gg r_s$, which does not significantly alter our conclusions.

\section{Surface conductivity}
\label{sec:surfaceconductivity}

We now turn our attention to the problem of how the 3D-distributed Coulomb impurities within the TI bulk affect the surface conductivity.  As discussed at the beginning of Sec.\ \ref{sec:surfacetheory}, we limit our consideration to the case where the Fermi level resides within the bulk band gap, where one can safely assume that all relevant bulk impurities are ionized.

Our primary result is an expression for the electron conductivity $\sigma$ of the surface as a function of the average 2D surface electron concentration $n$.  In particular, when $n \gg n_p$, where $n_p$ is the typical puddle concentration at $\mu = 0$ [see Eq.\ (\ref{eq:ne})], we find that the conductivity is given by
\be
\sigma \simeq \frac{e^2}{h} \frac{2 \sqrt{\pi}}{\alpha^2 \ln(1/\alpha)} \frac{n^{3/2}}{N},
\label{eq:sigman}
\ee
where $e^2/h$ is the conductance quantum.  At much smaller electron concentrations, $n \ll n_p$, the conductivity saturates at a value $\smin$, which we estimate as
\be 
\smin \simeq \frac{e^2}{h} \frac{1}{\pi \alpha \ln(1/\alpha)}.
\label{eq:smin}
\ee

To derive these results, we first note that in the limit of large chemical potential $\mu$, where the electron density is only weakly modulated by the disorder potential, one can show using the Boltzmann kinetic equation that for electrons with a massless Dirac spectrum the conductivity is given by \cite{Adam2007sct, Culcer2010tds, Culcer2008wms, DasSarma2011eti}
\be 
\sigma = \frac{e^2}{h} \frac{\mu \tau}{4 \hbar}.
\label{eq:sigmadef}
\ee 
Here $\tau$ is the momentum relaxation time.
In the limit of zero temperature, the scattering rate $1/\tau$ can be found by integrating the squared scattering potential produced by a given impurity over all impurities and over all scattering angles.  More simply, one can arrive at an expression for $1/\tau$ by taking the result for the scattering rate of a 2D layer of impurities with concentration $n_i$ at distance $z$ [for example, Eq.\ (38) of Ref.\ \onlinecite{Culcer2010tds}], replacing $n_i$ with $2 N dz$, and then integrating over all planes $z$  containing impurities.  This procedure gives
\be 
\frac{1}{\tau} = \frac{k_f \alpha \kappa_s}{4 \pi \hbar e^2} \int_0^\infty 2 N dz \int_0^\pi d \theta \left[ \phit (2 k_f \sin\frac{\theta}{2}; z ) \right]^2 (1 - \cos^2 \theta).
\label{eq:taudef}
\ee
In this equation, $k_f = \alpha \kappa_s \mu / e^2$ is the Fermi wavelength, $\phit (q; z) = (2 \pi e^2/\kappa_s q) \exp[-q z]/[1 + (qr_s)^{-1}]$ is the screened potential (in momentum space) created by a single impurity at position $z$, and $q = 2 k_f \sin (\theta/2)$ is the change in momentum associated with scattering by an angle $\theta$.  

Evaluating the integral of Eq.\ (\ref{eq:taudef}) at small $\alpha$ gives
\be 
\frac{1}{\tau} \simeq \pi \alpha \ln \left(1/\alpha \right) \frac{e^2 N}{\hbar \kappa_s k_f^2}.
\ee
Inserting this result for $\tau$ into Eq.\ (\ref{eq:sigmadef}) and substituting $\mu = e^2 k_f/\alpha \kappa_s$ and $k_f = \sqrt{4 \pi n}$ yields the result for conductivity announced at the beginning of the section, Eq.\ (\ref{eq:sigman}).

Equation (\ref{eq:sigman}) can be contrasted with the widely-used result for the 2D model of charge impurities \cite{Adam2007sct, DasSarma2011eti, Li2012tde, Culcer2010tds}, for which the conductivity is linearly proportional to the electron density: $\sigma/(e^2/h) \sim (1/\alpha^2)(n/n_i)$.  This difference can be understood conceptually by noting that, for large angle scattering, only those impurities at a distance smaller than the Fermi wavelength, $\lambda_f \sim n^{-1/2}$, contribute significantly to scattering.  One can therefore define, roughly speaking, an effective 2D concentration of scattering impurities as $N \lambda_f \sim N / n^{1/2}$.  Inserting $N/n^{1/2}$ for $n_i$ gives $\sigma \propto (1/\alpha^2)(n^{3/2}/N)$, similar to Eq.\ (\ref{eq:sigman}).  The remaining factor $1/\ln(1/\alpha)$ in Eq.\ (\ref{eq:sigman}) is related to low-angle scattering by distant impurities with $z \gg \lambda_f$.
 So far we are unaware of any transport data for TIs that shows $\sigma \propto n^{3/2}$.  Recent conductivity measurements on ultra-thin TIs (with thickness $\sim 10$\,nm $\ll \lambda_f$) suggest \cite{Kim2012sct} $\sigma \propto n$, consistent with the 2D model of impurities.

Our 3D model also produces a distinct result for the minimum conductivity $\smin$ that appears in the limit of small average electron concentration.  At small enough chemical potential that $\mu \ll e^2 N^{1/3}/\alpha^{2/3} \kappa_s$, the surface breaks into electron and hole puddles, and one can think that the effective carrier concentration saturates at $\sim n_p$ [see Eq.\ (\ref{eq:ne})].  An estimate of $\smin$ can therefore be obtained by setting $n \sim n_p$ in Eq.\ (\ref{eq:sigman}), which gives \cite{Skinner2013trp} the result of Eq.\ (\ref{eq:smin}).  2D models of disorder impurities also produce a minimum conductivity that is independent of the impurity concentration, but which has a different dependence on $\alpha$.  Specifically, at small $\alpha$ such models give \cite{Fogler2009npg, Adam2007sct} $\smin \sim (e^2/h) \ln(1/\alpha)$.  Our model suggests a minimum conductivity that is larger by a factor $\sim [\alpha \ln^2(1/\alpha) ]^{-1}$.

\section{TI Surface with a gap}
\label{sec:surfacegap}

\begin{figure}[htb!]
\centering
\includegraphics[width=0.48 \textwidth]{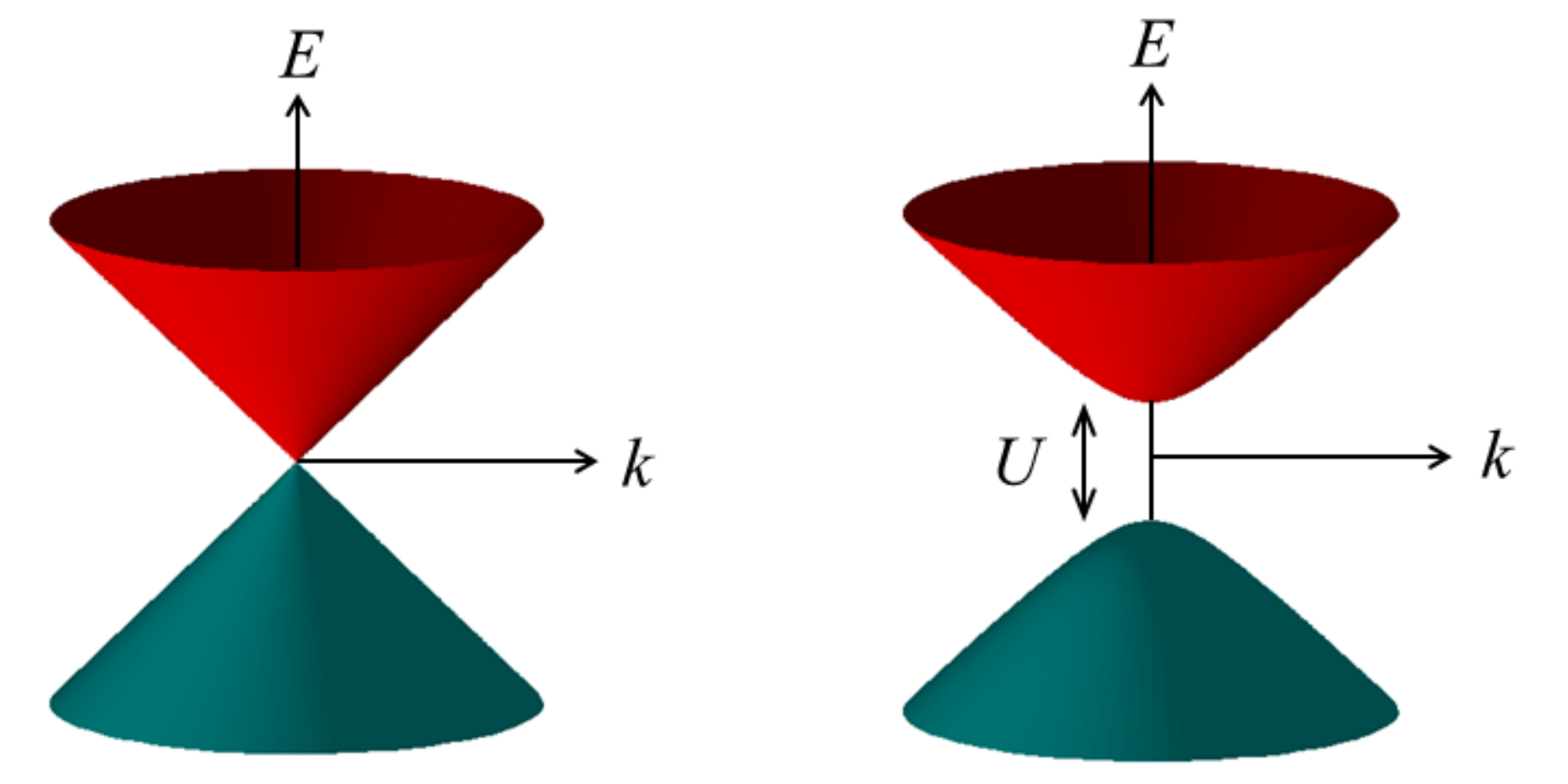}
\caption{(Color online)  Schematic illustration of a gap opening at the TI surface between the conduction band (upper, red) and valence band (lower, blue).  With the addition of some source of time-reversal symmetry breaking, the ungapped dispersion relation (left) acquires an energy gap $U$ (right).}
\label{fig:gap}
\end{figure}

In Secs.\ \ref{sec:surfacetheory} and \ref{sec:surfacetobulk}, we discussed the disorder potential created by Coulomb impurities at a gapless TI surface, whose massless spectrum is protected by time-reversal symmetry.  On the other hand, one can open a gap at the TI surface if one introduces some source of time-reversal symmetry breaking, such as an external magnetic field \cite{Fu2007tiw, Hanaguri2010mrl, Cheng2010lqt}, proximity to a magnetic material or magnetic impurities \cite{Chen2010mdf, Liu2009mis}, the proximity effect from an adjacent superconductor \cite{Fu2008spe}, or electron tunneling between two nearby TI surfaces \cite{Seradjeh2009eca, Zhang2010ctd} (see also the review of Ref.\ \onlinecite{Hasan2010c:t}).  The resulting gapped spectrum is illustrated schematically in Fig.\ \ref{fig:gap}. 

In this final section we briefly discuss how the presence of a gap with magnitude $U$ affects the disorder potential at the surface and the mid-gap density of states.  We focus our discussion around the case where the chemical potential $\mu = 0$, which roughly corresponds to the largest disorder potential and the minimum in the thermodynamic density of states.  Again, we limit our consideration to the case where the TI is sufficiently thick that one can describe impurities as three-dimensionally distributed.

In the absence of a gap, $U = 0$, the disorder potential is well-described by the results of the previous section.  In particular, the disorder potential width $\Gamma = \Gamma_0 \equiv (2^{1/6} \sqrt{\pi} /\alpha^{2/3}) (e^2 N^{1/3}/\kappa_s)$ 
[see Eq.\ (\ref{eq:gamma0})] and the average density of states $\tdos = \langle\nu \rangle_0 \equiv (\alpha^{4/3} /2^{1/3} \pi) (\kappa_s N^{1/3}/e^2)$
[see Eq.\ (\ref{eq:nu0})].  If the gap $U$ is small enough that $U \ll \Gamma_0$, then the disorder potential at the surface is essentially unaffected by the gap, since local fluctuations in the Fermi level are much larger than $U$.  For example, if the gapless surface spectrum is replaced with a ``massive" dispersion relation,
\be 
E = \pm \sqrt{(\hbar v k)^2 + (U/2)^2},
\label{eq:EU}
\ee 
as plotted in Fig.\ \ref{fig:gap}, then one can estimate the first-order effect of the gap by carrying out the same self-consistent procedure outlined in Sec.\ \ref{sec:surfacetheory}.  In particular, the gapped dispersion relation of Eq.\ (\ref{eq:EU}) has a corresponding density of states 
\be 
\nu(E; U) = \frac{|E|}{2 \pi \hbar^2 v^2} \Theta \left(|E| - U/2 \right),
\ee
where $\Theta(x)$ is the Heaviside step function.  At $U/\Gamma_0 \ll 1$, one can assume a Gaussian distribution of the Coulomb potential $\phi$ with some unknown variance $\Gamma^2$, integrate this distribution over $\phi$ multiplied by $\nu(-e\phi; U)$ to produce the thermodynamic density of states $\tdos$, and then use the self-consistency relation $\Gamma^2 = e^2 N /\kappa_s \tdos$ to arrive at a value for $\Gamma$ [see Eq.\ (\ref{eq:gammalargemu})].  Expanding the result of this procedure for small $U/\Gamma_0$ gives for the disorder potential width a slightly enhanced value
\be 
\Gamma(U) \simeq \Gamma_0 \left( 1 + \frac{U^2}{24 \Gamma_0^2} \right).
\ee 
Similarly, the thermodynamic density of states is slightly depleted:
\be 
\tdos \simeq \tdos_0 \left( 1 - \frac{U^2}{12 \Gamma_0^2} \right).
\ee 

On the other hand, if $U$ is much larger than $\Gamma_0$, then the surface screens poorly and the disorder potential grows.  In this case screening of the disorder potential by the surface happens only nonlinearly, through the formation of electron and hole puddles at locations where the magnitude of the Coulomb potential energy reaches the gap energy $U/2$.  This is similar to the bulk nonlinear screening discussed in Sec.\ \ref{sec:bulk}, and naturally produces $\Gamma(U) \sim U$.  The typical correlation length of the disorder potential at the surface (the nonlinear screening length) is given by
\be 
R_U \sim \frac{U^2 \kappa_s^2}{N e^4},
\ee 
as in Eq.\ (\ref{eq:Rg}), with $E_g \rightarrow U$.  

One can estimate the corresponding concentration of electrons/holes in surface puddles, $n_p$, by noting that a square area of size $R_U^2$ at the surface should contain enough electrons/holes to neutralize the net charge of Coulomb impurities in the adjacent cubic volume $R_U^3$ of the TI bulk.  This gives $n_p R_U^2 \sim \sqrt{N R_U^3}$, or in other words 
\be 
n_p \sim \frac{ e^2 N}{\kappa_s U}.
\label{eq:npU}
\ee 

The corresponding thermodynamic density of states can be estimated by noting that when the chemical potential $\mu$ is raised by an amount $\sim U/2$, surface hole puddles should dry up and be replaced by a correspondingly increased number of electron puddles.  This suggests $\tdos = d\mu/dn \sim n_p/U$, which gives
\be 
\tdos \sim \frac{e^2 N}{\kappa_s U^2}.
\label{eq:nuU}
\ee
Notice that if the gap $U$ is reduced to the point where $U \sim \Gamma_0$, then $\tdos \rightarrow \tdos_0$, as can be seen by comparing Eq.\ (\ref{eq:nuU}) with Eqs.\ (\ref{eq:gamma0}) and (\ref{eq:nu0}).

Of course, these estimates assume that the surface gap $U$ is smaller than the bulk band gap $E_g$, and consequently that $R_U \ll R$, so that impurities near the surface are not screened by bending of the bulk bands.  If the surface gap $U$ is larger than $E_g$, then the disorder potential variance is truncated at $\Gamma(U) \sim E_g$ due to bulk screening.  

We note that Eqs.\ (\ref{eq:npU}) and (\ref{eq:nuU}) were first derived in Ref.\ \onlinecite{Shklovskii1986soo} in the context of semiconductor heterostructures in a transverse magnetic field, where a 2D electron gas experiences disorder from adjacent 3D impurities and the gap $U$ in the kinetic energy spectrum is provided by the Landau level spacing $\hbar \omega_c$.  These authors also showed how the disorder potential is reduced and the density of states increased as the chemical potential $\mu$ is increased from zero \cite{Shklovskii1986soo}.  Specifically, $\Gamma \sim U - 2\mu$ and $\tdos \sim e^2 N /\kappa_s (U - 2\mu)^2$, provided that $U - 2\mu \gg e^2 N^{1/3}/\kappa_s$.  Of course, for TIs the effect of a transverse magnetic field goes beyond simply opening a single gap at the Dirac point \cite{Cheng2010lqt, Hanaguri2010mrl}.  We do not consider here the full problem of screening of Coulomb impurities in the presence of a magnetic field, but in principle this problem can be dealt with along the lines of Ref.\ \onlinecite{Shklovskii1986soo}.

\begin{acknowledgments}

The authors are grateful to Y. Ando, A. L. Efros, H. Beidenkopf, M. M. Fogler, M. S. Fuhrer, Yu. M. Galperin, J.\ Kakalios, Q. Li, M. M\"{u}ller,  N. P. Ong, and A.\ Yazdani for helpful discussions.  This work was supported primarily by the National Science Foundation through the University of Minnesota MRSEC under Award Number DMR-0819885.  T. Chen was partially supported by the FTPI.

\end{acknowledgments}

\bibliography{minireview}


\end{document}